\documentclass[prl,showpacs,twocolumn]{revtex4}
\usepackage{mathrsfs}
\usepackage{amsmath}
\usepackage{amssymb}
\usepackage{revsymb}
\usepackage{graphicx}
\usepackage{mathrsfs}

\begin{document}
\title{Relativistic ionization-rescattering with tailored laser pulses}
\author{Michael Klaiber$^{1,2}$}
\email{klaiber@mpi-hd.mpg.de}
\author{Karen Z. Hatsagortsyan$^1$}
\email{k.hatsagortsyan@mpi-hd.mpg.de}
\author{Christoph H. Keitel$^{1}$}
\email{keitel@mpi-hd.mpg.de}
\affiliation{$^1$Max-Planck Institut f\"ur Kernphysik, Saupfercheckweg 1, D-69117 Heidelberg, Germany\\
$^2$Theoretische Quantendynamik, Physikalisches Institut der Albert-Ludwigs-Universit\"at,
    Hermann-Herder-Stra\ss e 3, D-79104, Freiburg, Germany}

\date{\today}

\begin{abstract}

The interaction of relativistically strong tailored laser pulses with an
atomic system is considered.  Due to a special tailoring of the laser pulse,
the suppression of the relativistic drift of the ionized electron and a dramatic enhancement of the
rescattering probability is shown to be achievable. The high harmonic generation rate in
the relativistic regime is calculated  and shown to be
increased by several orders of magnitude compared to the case of
conventional laser pulses. The energies of the revisiting electron at the
atomic core can approach the MeV domain, thus rendering hard x-ray harmonics and nuclear reactions with single atoms feasible.

\end{abstract}

\maketitle
 
In intense laser-atom interaction phenomena the rescattering concept plays a central role \cite{tsm}.
Here the electron is ionized, propagated in the continuum by the laser
field, and finally driven back and scattered at the ionic core
in the case of above-threshold ionization (ATI)\cite{ATI}, ionizing further bound electrons in the case of non-sequential double ionization (NSDI) \cite{NSDI}, or recombining with the ionic core with  high-harmonic
generation (HHG) \cite{HHG}. It is highly desirable to increase the rescattering
electron energy \cite{review} as it can be employed to generate higher harmonics
\cite{ion}, to image attosecond dynamics of nuclear processes
\cite{circular}, or to initiate nuclear reactions with a single
atom/molecule \cite{mocken,muonic}. The electron energy increase
can not be achieved by a straightforward increase of the laser
intensity. When the laser intensity approaches the relativistic
regime, the laser magnetic field effect starts to play a role by
inducing a drift of the ionized electron in the laser propagation
direction which  severely suppresses the probability of the
electron revisiting the ionic core. There are several attempts to
circumvent this effect in order to increase the efficiency of
rescattering, particularly, using relativistic ions which
propagate in laser propagation direction \cite{mocken,ion}, or using two
counter-propagating laser beams with linear \cite{kylstra} or
equally handed circular polarization \cite{circular}, or generating harmonics via exotic positronium atoms in strong laser fields \cite{henrich}. On a different front, new techniques for generating attosecond pulses
have recently emerged based on HHG in gases \cite{as-train,as} and in plasmas interaction \cite{laser-plasma,laser-foil}. The usage of attosecond pulse trains (APT's) to enhance HHG in a strong laser field has been shown in \cite{keller}.

\begin{figure}
  \begin{center}
    \includegraphics[width=0.4\textwidth,clip=true]{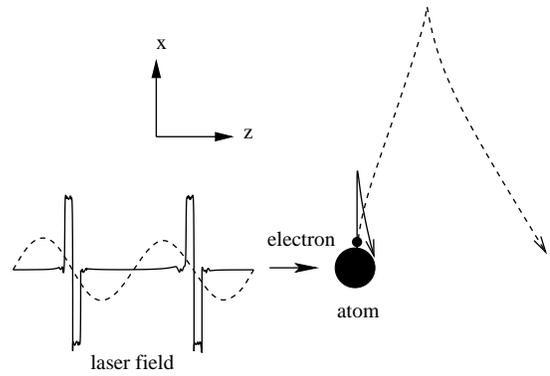}
\caption{By using optimized tailored pulses in the form of an APT instead of conventional
sinusoidal pulses it is possible to reduce  significantly the drift of the ionized electrons
due to the Lorentz force of the laser field (the ionized electron trajectory in the sinusoidal pulse is shown as a dashed line, in the tailored pulse as a solid line). Therefore, the rescattering probability can be increased by
several orders of magnitude in the relativistic regime of
laser-atom interactions. $x$ and $z$ are the laser polarization
and propagation directions, respectively.}
    \label{Trajec}
  \end{center}
\end{figure}

\begin{figure}
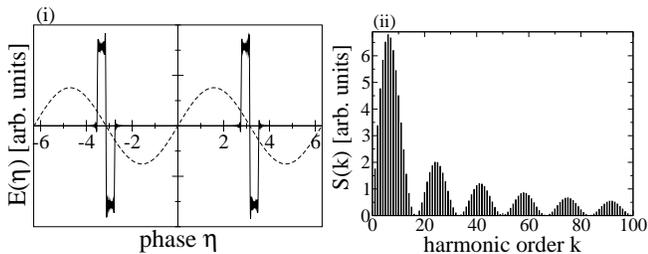

  \begin{center}
\includegraphics[width=4.2cm]{klaiberfig2i.eps}
\includegraphics[width=4.2cm,clip=true]{klaiberfig2ii.eps}
\caption{The tailored laser pulse in the form of an APT, with each pulse duration in the train being $\tau=0.12 \pi$ a.u.: (i) Phase dependence of the electric field ($E(\eta)$) of the tailored pulse (solid line) compared with a sinusoidal field (dashed line) with equal average intensities and central angular frequency $\omega =0.1$ a.u.; (ii) Frequency spectrum ($S(k)$) of the tailored pulse. The frequency components are phase locked.}
    \label{pulse}
  \end{center}
\end{figure}

\begin{figure}[b]
  \begin{center}
    \includegraphics[width=4cm]{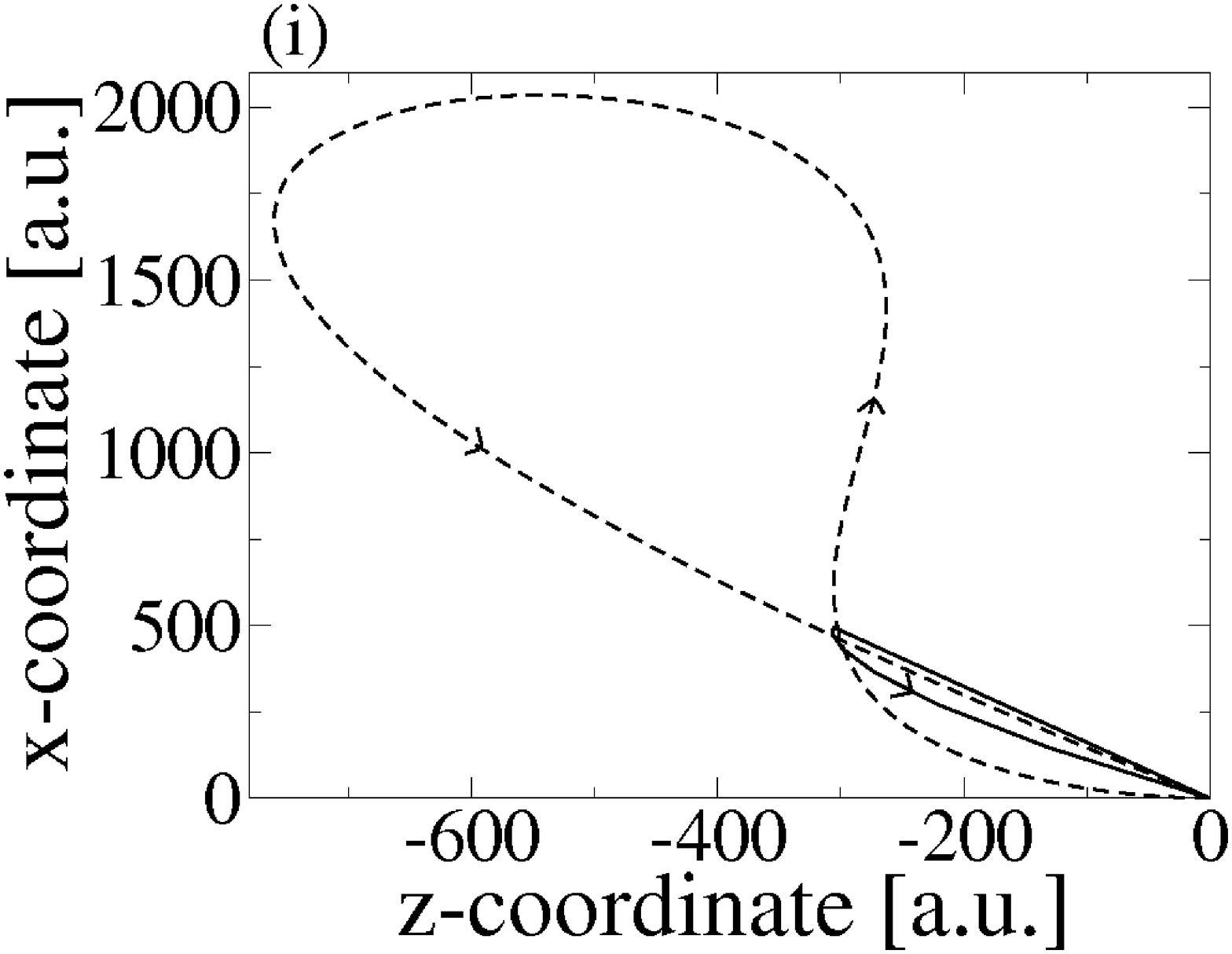}
    \includegraphics[width=4cm]{klaiberfig3ii.eps}
    \caption{Trajectory and kinetic energy of an HHG  electron in the tailored wave (solid line)
      and in a sinusoidal wave (dashed line) with an averaged laser intensity of $5\times10^{19}$ W/cm$^2$. ($x$ and $z$ are the polarization and
      propagation direction of the laser field, respectively.) The electron is ionized at a phase of -1.24 in the sinus wave and recombines at a phase of 2.68. In the
    tailored wave the electron is ionized at a phase of -2.78 and recombines at a phase of 3.14.}
    \label{tra}
  \end{center}
\end{figure}

In this letter we show how efficient recollisions in the relativistic regime are feasible by employing
specially tailored strong laser pulses in the form of APT's (see Fig.\ref{Trajec}).   
As an indicator of strong field recollision phenomena we
consider the HHG process from an ion in a relativistically strong tailored laser pulse (see Fig.\ref{pulse}). While in \cite{keller} the  weak APT serves  to solely control the initial conditions of the ionized electron in the strong laser field, we consider HHG in  relativistically strong APT.
The temporal tailoring of the laser pulse is aimed to concentrate the ionizing and afterwards the
accelerating laser forces in short time intervals within the laser
period (maintaining  the average intensity of the pulse constant) which substantially reduces the relativistic drift.  This results from the fact that in the tailored laser pulse, fragments in the electron trajectory are avoided, in contrast to the sinusoidal laser pulse, in which the electron acceleration is compensated by deceleration without a net energy gain by the electron, but during which nevertheless the electron would continue to drift in the laser propagation direction (see Fig.~\ref{tra}(ii)). This substantially decreases the time span during which the electron moves with a relativistic velocity. The latter results in a shorter drift in the laser propagation direction, and therefore, in an increase of the recombination probability in the case of the tailored pulse.

The scenario of the electron motion in the tailored laser pulse looks as follows.
The trajectory of the rescattering electron  begins at the end of the period with maximal field strength of the laser pulse, shortly before the field free region. The initial spike of the laser pulse ionizes the electron and initiates its excursion with a very low initial velocity. The acceleration of the electron takes place during the next spike after the field free region. One can see from Fig.~\ref{tra} that in the tailored wave the electron gains most of its energy only shortly before
the recombination and the trajectory of the electron in the tailored pulse is spatially significantly shorter than in the sinusoidal wave.

The procedure for tailoring the laser pulse is the following.  The laser pulse with temporal form as shown in Fig.\ref{pulse} has been chosen with the additional condition of vanishing the
field average of the pulse  $<E>=0$. Then, the Fourier-expansion of this pulse is used via $K=100$ harmonic terms to optimize the field expansion coefficients by maximizing the rescattering (HHG) probability. The optimized tailored pulse has the form of an APT which includes three key features. The first is a large ionization peak with a strong decay after the point of ionization (the stronger decay the larger HHG efficiency). This follows by a long region with an only weak field strength (the longer the better) which is resumed by a short and strong pulse that induces most of the energy of the ionized electron. The optimized pulse along with its spectrum is shown in Fig.~\ref{pulse}. The rescattering rate resulting from this pulse can be enhanced by using more frequency components in the APT. In the calculated spectra that will follow, we employ 100 frequency components. When using only $K=30$ harmonics in the tailored pulse, the HHG yield decreases by approximately 2 orders of magnitude at laser intensity of $5\times 10^{18}$ W/cm$^2$ and by 5 orders of magnitude for $5\times 10^{19}$ W/cm$^2$. 
A possible experimental way for the above mentioned tailoring of the laser pulses is to employ relativistic harmonics generated during a laser-foil interaction in the so-called sliding mirror regime \cite{laser-foil} which yields very strong APT's with a conversion efficiency of a few percent. Additionally the technique for the spectral filtering \cite{as-train} should be applied to form the necessary spectral dips as shown in Fig.~\ref{pulse}. Significantly less  spectral filtering would be necessary when using $K=30$ harmonics and a APT spectrum with one hump, though in this case the HHG yield will be reduced as indicated above.

Given the tailored laser pulse, we proceed to calculate the HHG
intensity. Our investigation of HHG in an optimized tailored laser pulse is based
on the solution of the Klein-Gordon equation in the strong field
approximation (SFA) \cite{sfa}. 
We consider the HHG process of an atomic system by a linearly
polarized plane laser field for the
relativistic parameter regime. The direction of the laser field polarization is along the $x$ axis, that of the magnetic
field along the $y$, and of the laser propagation along the $z$ axis. The amplitude of HHG within the SFA based on the
Klein-Gordon equation in the single-active
electron approximation is given by the following expression
\cite{rel_hhg} (atomic units are used throughout the paper):
\begin{eqnarray}
  M_{n}&=&
    -i \int
    d^4x^{\prime}\int d^4x^{\prime\prime} \left\{\Phi(x^{\prime})^*
    \right. \nonumber \\
      &\times &   \left. V_{H}(x^{\prime})G^V(x^{\prime},x^{\prime\prime})
     V_{A}(x^{\prime\prime})\Phi(x^{\prime\prime})\right\}
     \label{mhhg}
\end{eqnarray}
with $n$ the harmonic order, $\Phi$ the bound state wave function,
$V_H(x)=2\mathbf{A}_H(x)\cdot(\hat{\mathbf{p}}+\mathbf{A}(\eta)/c)$,
$V_{A}(x)=2iV(x)/c^2\partial_{t}+V(x)^2/c^2$, $V(x)$ the atomic
potential, $c$ the speed of light, $\hat{\mathbf{p}}$ the momentum
operator of the electron, $\mathbf{A}(\eta)$ the vector potential
of the laser field in the radiation gauge, $k^{\mu }=(\omega
/c,0,0,k)$ its 4-wave-vector, $\eta =k^{\mu}x_{\mu}$, $x^{\mu}=(ct, \mathbf{x})$
the time-space coordinate,  $\mathbf{A}_H(x)$ the matrix element
of the vector potential of the high harmonic field in the second
quantization for an one photon emission process, and $G^V(x^{\prime},x^{\prime\prime})$
the Klein-Gordon Volkov Green function given in
Ref.~\cite{rel_hhg}. $\Phi$ is an eigenstate of the physical
energy operator in the radiation gauge and can be
approximated as $\Phi(x)=\phi_0(\mathbf{x})\exp\left\{-i[(c^2-I_p)t+ \mathbf{x}\cdot\mathbf{A}/c]\right\}/\sqrt{2(c^2-I_p)}$
\cite{klaiber1}, with the nonrelativistic ground state wave function
$\phi_0(\mathbf{x})$. Introducing the integration variable transformations
$\eta^{\prime}=k^{\mu}x_{\mu}^{\prime}$ and $\eta^{\prime \prime}
=k^{\mu}x_{\mu}^{\prime \prime}$ and taking into account that the
atomic wave function characteristic length is much smaller than
the laser wavelength, the amplitude of HHG can be written:
\begin{eqnarray}
  M_n=\int^
  {\infty}_{-\infty}d\eta^{\prime}\int^{\eta^{\prime}}_{-\infty}d\eta^{\prime \prime}\int d^3\mathbf{q}\,
  m^H(\mathbf{q},\eta^{\prime},\eta^{\prime \prime}) \nonumber \\
  \times\exp\left\{-i\left[S(\mathbf{q},\eta^{\prime},\eta^{\prime
  \prime})-n\eta^{\prime}\right]\right\}
    \label{mhhg2}
\end{eqnarray}
with the HHG matrix element for emission in laser polarization direction
\begin{eqnarray}
  m^H(\mathbf{p},\eta^{\prime},\eta^{\prime \prime})&=&-\frac{c^2
  (p_x+A(\eta^{\prime})/c)}{\varepsilon_\mathbf{p}\omega^2} \label{mh}\\
  &\times & \tilde{\phi}^*_0\left(\mathbf{p}+\frac{\mathbf{A}(\eta^{\prime})}{c}-\frac{{\bf k}}{\omega}(\varepsilon_{{\bf p}}+I_p-c^2)\right)
  \nonumber\\
  &\times &\left<\mathbf{p}+\frac{\mathbf{A}(\eta^{\prime})}{c}-\frac{{\bf k}}{\omega}(\varepsilon_{{\bf
  p}}+I_p-c^2)\left|V\right|0\right>,
  \nonumber
\end{eqnarray}
with the Fourier transform of the ground state wave function
$\tilde{\phi}_0$ and
$\mathbf{\varepsilon}_\mathbf{p}=\sqrt{c^4+c^2\mathbf{p}^2}$.
$S(\mathbf{p},\eta,\eta^{\prime})= \int^{\eta}_{\eta^{\prime}}
d\tilde{\eta}\left(\tilde{\varepsilon}_{\mathbf{p}}
  (\tilde{\eta})-c^2+I_p\right)/\omega$
is the quasi-classical action
and the relativistic energy is given by
\begin{eqnarray}
  \tilde{\varepsilon}_{\mathbf{p}}(\eta)=\varepsilon_{\mathbf{p}}
  +\frac{\omega}{k\cdot p}
  \left(\mathbf{p}+\mathbf{A}(\eta)/2c\right)\cdot\mathbf{A}(\eta)/c.
\end{eqnarray}
The differential rate of HHG is then given by
\begin{eqnarray}
  \frac{dw_n}{d\Omega}=\frac{\omega^2}{(2\pi)^3}\frac{n\omega}{c^3}|M_n|^2
  \label{rate}
\end{eqnarray}
with the solid angle of emission $\Omega$.

\begin{figure}
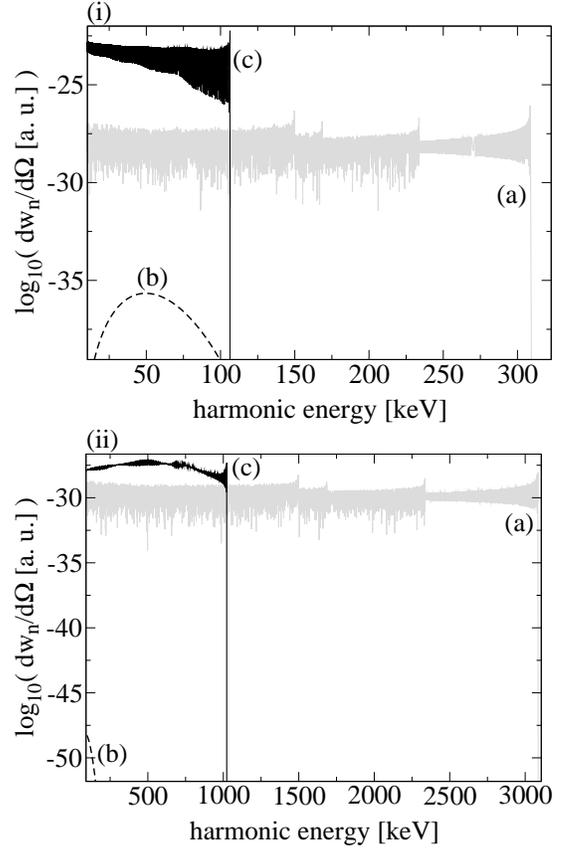

  \begin{center}
    \includegraphics[width=0.4\textwidth]{klaiberfig4i.eps}
    \includegraphics[width=0.4\textwidth,clip=true]{klaiberfig4ii.eps}
         \caption{Harmonic emission rate in laser polarization direction via $\log_{10}(dw_n/d\Omega)$ in Eq.~(\ref{rate}),
      as function of the harmonic energy with a laser intensity of (i) $5\times 10^{18}$ W/cm$^2$
      and (ii) $5\times 10^{19}$ W/cm$^2$. The  main angular frequency
      is $\omega=0.1$ a.u.. The ionization potential is (i) $I_p=28$ a.u. and (ii) $I_p=62$ a.u.,
      respectively: (a, grey) within the dipole approximation and a sinusoidal field,
      (b, dashed) with respect to the Klein-Gordon
      equation and a sinusoidal field, (c, black) with respect to the Klein-Gordon
      equation and a tailored pulse.}
    \label{h}
  \end{center}
\end{figure}

Using the low-frequency approximation $K\omega \ll I_p$ with the
ionization potential $I_p$, the integral in Eq.(\ref{mhhg2}) can
be calculated via the saddle-point method. The
saddle point conditions along with the cutoff condition determine
the ionization and recombination phases
$\tilde{\eta}^{\prime\prime},\tilde{\eta}^{\prime }$, as well as
the cutoff harmonic number $\tilde{n}$:
\begin{eqnarray}
 \tilde{\varepsilon}_{\tilde{\mathbf{q}}}(\eta^{\prime})&=&c^2-I_p+n\omega
  \nonumber\\
  \tilde{\varepsilon}_{\tilde{\mathbf{q}}}(\eta^{\prime \prime})&=&c^2-I_p
  \nonumber\\
  \frac{d\tilde{\varepsilon}_{\tilde{\mathbf{q}}}(\eta^{\prime})}{d \eta^{\prime
  \prime}}&=&0,
\end{eqnarray}
with the drift momentum of the ionized electron
$\tilde{q}_x = -\frac{\int^{\tilde{\eta}'}_{\tilde{\eta}''}d\tilde{\eta}
A(\tilde{\eta})/c}{\tilde{\eta}'-\tilde{\eta}''}$, $\tilde{q}_y=0$,   $\tilde{q}_z=\frac{\tilde{q}_x^2+q_m^2/2}{\sqrt{c^2-q_m^2-\tilde{q}_x^2}}$,
with  $q_m^2=-(1/c^2)\int^{\tilde{\eta}'}_{\tilde{\eta}''}d\tilde{\eta}
\mathbf{A}(\tilde{\eta})^2/(\tilde{\eta}'-\tilde{\eta}'')$~\cite{rel_hhg}. In tailoring the laser
pulse, we fix the harmonic cutoff $\tilde{n}$ and by changing the
shape of the pulse, minimize the imaginary part of the ionization
time Im$\{\tilde{\eta}^{\prime \prime}\}$ which determines the
plateau height and, therefore, the rescattering efficiency.
Further, the atomic potential $V(x)$ is approximated by
a zero-range potential \cite{zrp}. The ionization potential
$I_p$ is adapted to the threshold of the barrier suppression
ionization model.

The HHG spectrum with the tailored
pulse is compared with two calculations for the harmonic emission
using a sinusoidal field. While the first calculation is based on the dipole approximation (DA), the second one is fully-relativistic. We consider two
regimes: the moderately relativistic regime with a laser intensity of the tailored pulse
$5\times 10^{18}$ W/cm$^2$ and the strongly relativistic regime with a laser
intensity of $5\times 10^{19}$ W/cm$^2$, each with a laser ground angular frequency of 0.1 a.u.,
corresponding to a laser wavelength of $\lambda=456$ nm. Taking into account a few percent efficiency in the APT generation via laser-foil interaction \cite{laser-foil}, as well as about $20\%$ intensity reduction because of  spectral filtering in the case of $K=100$ spectral components in APT, the initial laser intensity before tailoring should be three orders of magnitude larger. For pulses involving fewer high harmonics this loss may be reduced.

In the moderately relativistic regime (see Fig.\ref{h}(i)), the
harmonic emission rate with conventional laser field is
suppressed by over seven  orders of magnitude in the relativistic treatment
compared with the results in the DA. Whereas the
tailored pulse yields a harmonic emission rate that is nearly four
orders of magnitude higher than the rate in the dipole
approximation. Further the spectrum induced by the tailored pulse as well as the spectrum induced by  the sinusoidal field using the DA
 is highly
oscillating, resulting from interference of at least two
trajectories of the ionized electron. In the case of the tailored
pulse, there are exactly two, since the trajectories with multiple return to the ionic
core are very unlikely.

The HHG spectrum in the strong relativistic regime  with a
laser intensity of $5\times 10^{19}$ W/cm$^2$ is shown in Fig.~
\ref{h}(ii). 
Whereas the tailored pulse spectrum is still
 more intense than the spectrum in the DA with a conventional laser
field, the relativistic treatment of HHG with sinusoidal fields yields
a rate that is negligibly small. From the HHG spectrum one can see that
cutoff energies of more than 1 MeV are reachable with the
probability not less than in the DA.

The enhancement of the HHG rate via tailored pulses compared with the one via a sinusoidal field is due to the suppressed relativistic drift. This can be deduced by comparing
the trajectory of the ionized cutoff electron driven by the tailored pulse with one driven by
the sinusoidal pulse, see Fig.~\ref{tra}(i). The excursion of the
ionized electron in polarization and propagation direction in the
tailored pulse is shorter, though the averaged intensity of the pulses are the same. Moreover, the HHG emission angle with respect to the laser polarization direction in the tailored pulse is significantly smaller than in the conventional sinusoidal field (see Fig.~\ref{tra}(i)).
In other terms, the suppressed drift can be deduced by the value of initial electron velocity in the laser propagation direction, necessary for the electron rescattering in the relativistic regime. In the tailored pulse it is much smaller compared to that for a sinusoidal pulse. In fact, in the strongly relativistic regime, the numerical values for the initial velocity in the tailored pulse is -6 a.u., whereas for the sinusoidal field -97 a.u.. The initial velocity of the tunneled electron determines the exponential damping factor of the electron tunneling probability, yielding a higher ionization rate and, therefore, a higher HHG efficiency, in the tailored pulse as compared to the sinusoidal field.

The reason for the increased
HHG rate with the tailored laser pulse over the dipole
approximation rate is that the laser electric field strength at
the time of ionization is higher in the tailored pulse at the same
average laser intensity. The numerical values are
$|E(\tilde{\eta}^{\prime\prime})|=$ 74 a.u. for the tailored
pulse and $|E(\tilde{\eta}^{\prime\prime})|=36$ a.u. for the
sinusoidal field for a laser intensity of $5\times 10^{19}$
W/cm$^2$.

Further, one can make a statement about the stability of the HHG process driven by such a tailored laser pulse.  A random variation of the amplitude of the frequency components by 5 percent, and of the phases within the range not exceeding $\pi/100$, results in a variation of the HHG yield within one order of magnitude.

Concluding, employing specially tailored laser pulses in the form of an APT allows to suppress the relativistic drift of the ionized electron and
to realize ionization-rescattering in the relativistic regime.
HHG in the hard x-ray domain and the initiation of nuclear reactions with single ions thus become feasible.

Funding by Deutsche Forschungsgesellschaft via KE-721-1 is acknowledged.

\end{document}